\documentclass[12pt]{article}
\usepackage{fullpage}

\usepackage{threeparttable}
\usepackage{lscape}

\usepackage{graphicx}
\usepackage{caption}
\usepackage{subcaption}

\usepackage{wasysym}
\usepackage{longtable}
\usepackage[none]{hyphenat}
 \usepackage{setspace}
 \usepackage{url}

\usepackage{lineno}
\usepackage{epstopdf}
\usepackage{caption}

\newcommand{\Dmel}{\emph{D. melanogaster}}

\newcommand{\Dyak}{\emph{D. yakuba}}
\newcommand{\Dsant}{\emph{D. santomea}}

\newcommand{\Dsech}{\emph{D. sechellia}}

\newcommand{\cis}{\emph{cis}}
\newcommand{\trans}{\emph{trans}}

\newcommand{\Dros}{\emph{Drosophila}}

\newcommand{\Rebekah}{Rebek\mbox{}ah }

\makeatletter
\renewcommand{\@biblabel}[1]{\quad#1.}
\makeatother

\usepackage{textcomp}

\begin{document}

\author{\Rebekah L. Rogers$^{1*}$, Cathy C. Moore$^1$, Nicholas B. Stewart$^{1,2}$}


\title{New gene formation in hybrid \Dros}
\date{}

\maketitle
\noindent1.  Dept of Bioinformatics and Genomics, University of North Carolina, Charlotte, NC \\
\noindent2.  Dept of Biology, Ft. Hays State University, Hays, KS \\

\noindent{}Keywords: New gene formation, Hybrid \Dros, Allele-specific expression

\noindent*  Corresponding Author: Rebekah.Rogers@uncc.edu \\
\clearpage
\subsection*{Abstract} 
The origin of new genes is among the most fundamental processes underlying genetic innovation.  The substrate of new genetic material available defines the outcomes of evolutionary processes in nature.   Historically, the field of genetic novelty has commonly invoked new mutations at the DNA level to explain the ways that new genes might originate.  In this work, we explore a fundamentally different source of epistatic interactions that can create new gene sequences in hybrids.  We observe ``bursts'' of new gene creation in F$_1$ hybrids of \Dyak {} and \Dsant, a species complex known to hybridize in nature.  The number of new genes is higher in the gonads than soma.   We observe asymmetry in new gene creation based on the direction of the cross.  Greater numbers of new transcripts form in the testes of F$_1$ male offspring in \Dsant {} $\female \times$  \Dyak {} $\male$ crosses and greater numbers of new transcripts appear in ovaries of F$_1$ female offspring of \Dyak {} $\female \times$ \Dsant {} $\male$.  These \emph{loci} represent wholly new transcripts expressed in hybrids, but not in either parental reference strain of the cross.  We further observe allelic activation, where transcripts silenced in one lineage are activated by the transcriptional machinery of the other genome.  These results point to a fundamentally new model of new gene creation that does not rely on the formation of new mutations in the DNA.   These results suggest that  bursts of genetic novelty can appear in response to hybridization or introgression in a single generation.  Ultimately these processes are expected to contribute to the substrate of genetic novelty available in nature, with broad impacts on our understanding new gene formation and on hybrid phenotypes in nature. 

\clearpage
\subsection*{Introduction}
  The ways that cellular mechanisms produce new genetic material remains a fundamental question in evolutionary theory. As organism adapt to changing environments, new genes are required for genomes to respond to selective pressures.  Moreover, the genetic basis of population variation in nature ultimately rests on the substrate of genetic novelty that appears.  Duplication has long been held as a source of genetic novelty that is important for adaptation and innovation \cite{Ohno,Conant2008}.   Genes with new roles can also appear when genome sequences are rearranged, if they acquire novel expression profiles \cite{de2009, bhutkar2007, harewood2014} or create chimeric constructs that drive transcription in wholly new locations \cite{rogers2015N, stewart2019}.  A rising interest has come as new high throughput assays have allowed us to study the origins of \emph{de novo} genes that appear from previously untranscribed regions \cite{mclysaght2016, schlotterer2015, Zhao2014,carvunis2012}.  New work has attempted to explain how \emph{de novo} genes might arise and then evolve to form new open reading frames \cite{carvunis2012,siepel2009,schlotterer2015}.  Most models invoke some means of novel transcription, often though \cis {} regulatory evolution followed by formation of new translation start sequences \cite{Zhao2014,carvunis2012,siepel2009,schlotterer2015}. However, partial gene duplication and rearrangement can form both transcription and translation start sites in new genes \cite{stewart2019, rogers2017}.  

These models of new gene formation are most often discussed in the context of mutation, that is changes in DNA underlie genetic changes producing new transcripts.   However, another possibility arises whereby new genes may be created without any underlying alteration to DNA sequence.   During hybridization abnormal regulatory changes may produce transcription patterns not observed in any parental species \cite{Landry2007, wittkopp2004, Ranz2004, emerson2010,Coolon2014}.  Regulatory changes in hybrids have long been studied in the context of regulatory evolution for existing genes and their contribution to species differences \cite{Coolon2014, wittkopp2004,emerson2010, landry2005}.  However, hybrid systems have not been fully explored in terms of their potential to create entirely new genes that may serve as a substrate that might facilitate evolutionary change.   

Early array-based assays were sufficient to study gene expression changes for known open reading frames \cite{Ranz2004, wittkopp2004}.   As high throughput whole transcriptome RNA sequencing has progressed, it is possible to identify more unusual changes to transcripts and expression profiles without the need to identify probes or annotate sequences prior to data acquisition \cite{emerson2010,Coolon2014}.  This technological advance allows us to survey new genes that may appear in hybrid systems, independently from annotations in either parental species.  This new technical landscape makes it possible to identify new gene formation in addition to expression changes for existing annotations.

Early studies of hybrid gene expression most often used artificial lab generated hybrids for model organisms that are unlikely to occur in nature \cite{Ranz2004,wittkopp2004}.  However, as genetic and genomic resources have improved outside the standard models, it is possible to directly assay regulatory phenotypes in species that are known to experience introgression \cite{Llopart2012}.  \Dyak {} and \Dsant {} are known to hybridize in nature \cite{Llopart2012,Turissini2017, Llopart2005x, Llopart2005, Lachaise2000}, raising the potential for hybridization to influence downstream evolutionary processes.  Determining the full spectrum of genetic novelty in hybrids represents a first step to understanding how new transcripts and expression profiles may shape variation in ways that are directly related to impacts of introgression in nature.  

To address genetic novelty in hybrid \Dros, we acquired RNAseq data for male and female germline and soma in the parental species and hybrid offspring of \Dyak {} and \Dsant.  We identify new genes formed in F$_1$ hybrid \Dros, where the number of newly expressed \emph{loci} per sex depends on the direction of the cross. We observe a greater number of new genes in males when \Dyak {} donates the paternal genotype, and a greater number of new genes in females when the maternal genotype is \Dyak.  We further identify allele specific activation, where a \emph{locus} previously silent in one species becomes reactivated in the F$_1$.  

Together these results suggest a rich array of new gene formation and allelic activation in hybrid \Dros, that can influence downstream evolution when species exchange genetic variation through introgression. 

\subsection*{Results}
\subsubsection*{New gene formation}

\begin{figure}

            \includegraphics[width=.5\textwidth]{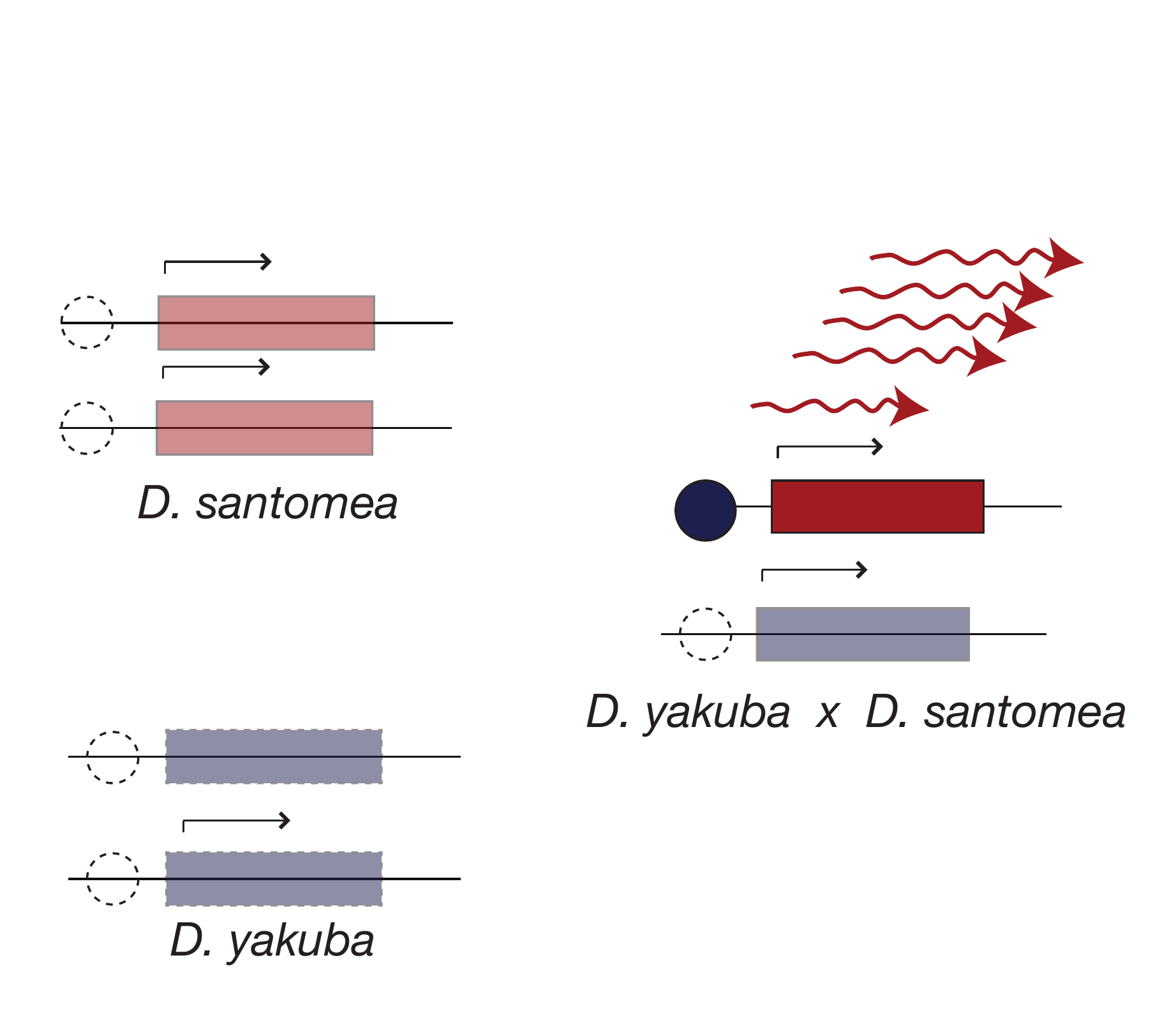}
            \caption{\label{NewGeneMechanism} New gene formation via epistatic interactions of \cis {} and \trans {} factors in hybrids.  A region of the genome has potential to drive expression with a \trans {} factor that is not present in the parental strain.  Hybridization brings an active \trans {} factor to this previously unusued \cis {} regulatory domain.  Transcription is driven from a locus where no expression occurred before.}
\end{figure}

 Hybrid genomes contain alternative genetic backgrounds not typically seen in non-hybrid individuals.  In such crosses, genetic background can influence coding potential at existing \emph{loci} in the DNA, even when there has been no underlying new mutation.  Through the combination of pre-existing pre-primed \cis {} regulatory modules and \trans {} acting factors, it is possible to form new genetic material (Figure \ref{NewGeneMechanism}). 
 
\begin{figure}
    \centering
    \begin{subfigure}[b]{0.3\textwidth}
        \includegraphics[width=\textwidth]{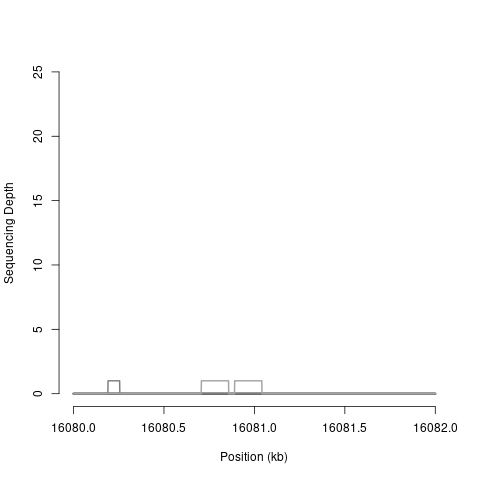}
           \end{subfigure}
    ~ 
    \begin{subfigure}[b]{0.3\textwidth}
        \includegraphics[width=\textwidth]{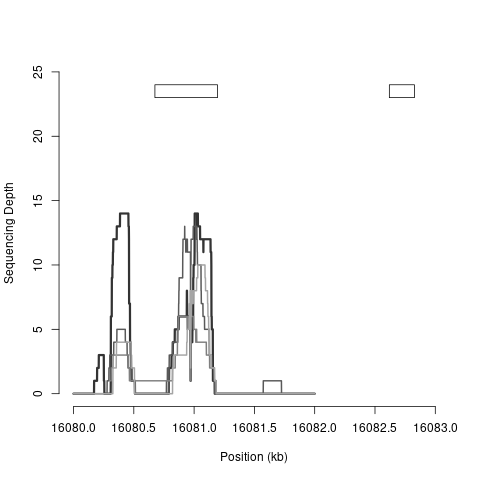}
          \end{subfigure}
    ~ 
    \begin{subfigure}[b]{0.3\textwidth}
        \includegraphics[width=\textwidth]{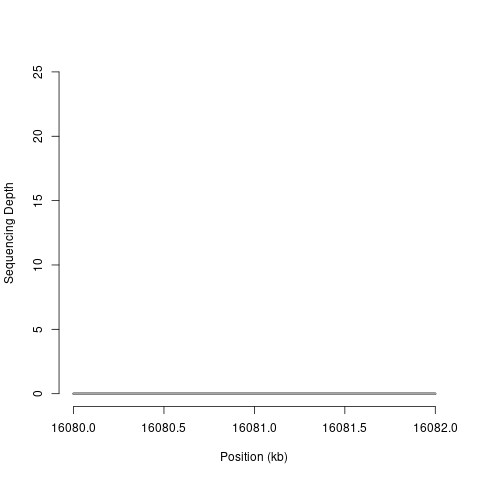}
      
    \end{subfigure}
    \caption{\label{XLOCEx} New gene formation on the X chromosome at 16,080,000-16,082,000 bp.   Plots show RNAseq coverage from testes plus accessory glands in 4 replicates each from A) \Dyak {} reference strains B)Adult male progeny from \Dyak {} $\male \times$ \Dsant {} $\female$ C) \Dsant {} reference strains.  The sequence does not match any annotated gene in \Dyak {} or \Dmel.  It does not show sequence similarity with repetitive elements nor does it align to more than one region in the \Dyak {} reference assembly. This \emph{locus} shows no expression in any other tissue or cross included in the study.  The \emph{locus} is unique, not repetitive, nor does it match any annotated gene sequence in \Dmel.  Two exons form in this region, which are identified as independent genes in cuffdiff, due to no read-pairs suggesting fused stranscripts.}
\end{figure}

We have observed evidence of new gene formation in \Dyak-\Dsant {} hybrids (Figure \ref{NewGeneMechanism}-\ref{XLOCEx}, Table \ref{NewGenes}).    We identify 31 \emph{loci} with no pre-existing gene annotation that are expressed in hybrids, but with no expression in either parental species.   A total of 18 of these are expressed in the testes of adult males in \Dsant {} $\female \times$  \Dyak {} $\male$ crosses, while 10 appear with expression in the ovaries in \Dyak {} $\female \times$ \Dsant {} $\male$ crosses (Table \ref{NewGenes}).  The asymmetry across tissues, sexes, and cross type is beyond null expectations (Table \ref{NewGenes},  $P=1.488\times 10^{-6}$).  The observed asymmetry with respect to crosses is consistent with phenotypic and molecular effects of the X in speciation and the bias toward damage in the heterogametic sex \cite{presgraves2018}.  

Genes are located on autosomes, the X, and on smaller unlocalized assembly contigs.   Testes genes lengths are 200-15,000 bp of genomic sequence from transcription start to stop, though all but 3 are less than 1000bp long.   New genes identified in ovaries span 218-1079 bp. Long reads validate 2/2 testes genes, and 6 of the ovaries new genes out of 10, for an overall confirmation rate of 75\%. Isoform long read sequencing uses size selection that may be biased against the smallest transcripts, reducing confirmation rates modestly. All appear tissue specific.   Reference genome bias can create technical artifacts in RNAseq analysis.  To ensure that results were not solely driven by the \Dyak {} reference sequence, we repeated analysis using the new \Dyak {} Tai18E2 contiguous PacBio reference genome (NCBI GCA\_016746365).  We identify 13 new genes formed in the testes and 4 new genes formed in ovaries.  These new genes are identified on the X and the autosomes, with no clear patterns of enrichment by chromosome.  Hence, results suggest that evidence for new gene formation in hybrids is robust to the choice of reference genome.

In addition to new genes that form, we observe 24 previously annotated gene sequences that follow the same expression profile with novel activation in hybrids (Table \ref{ActivatedGenes}).  While these 24 \emph{loci} are silent in both parental species, they may be expressed in reference strains at other times or in other tissues. These new regulatory profiles for existing genes are a source of genetic changes that may contribute to abnormal phenotypes in hybrids. 

\subsubsection*{Allelic activation }
\begin{figure}

        \includegraphics[width=.5\textwidth]{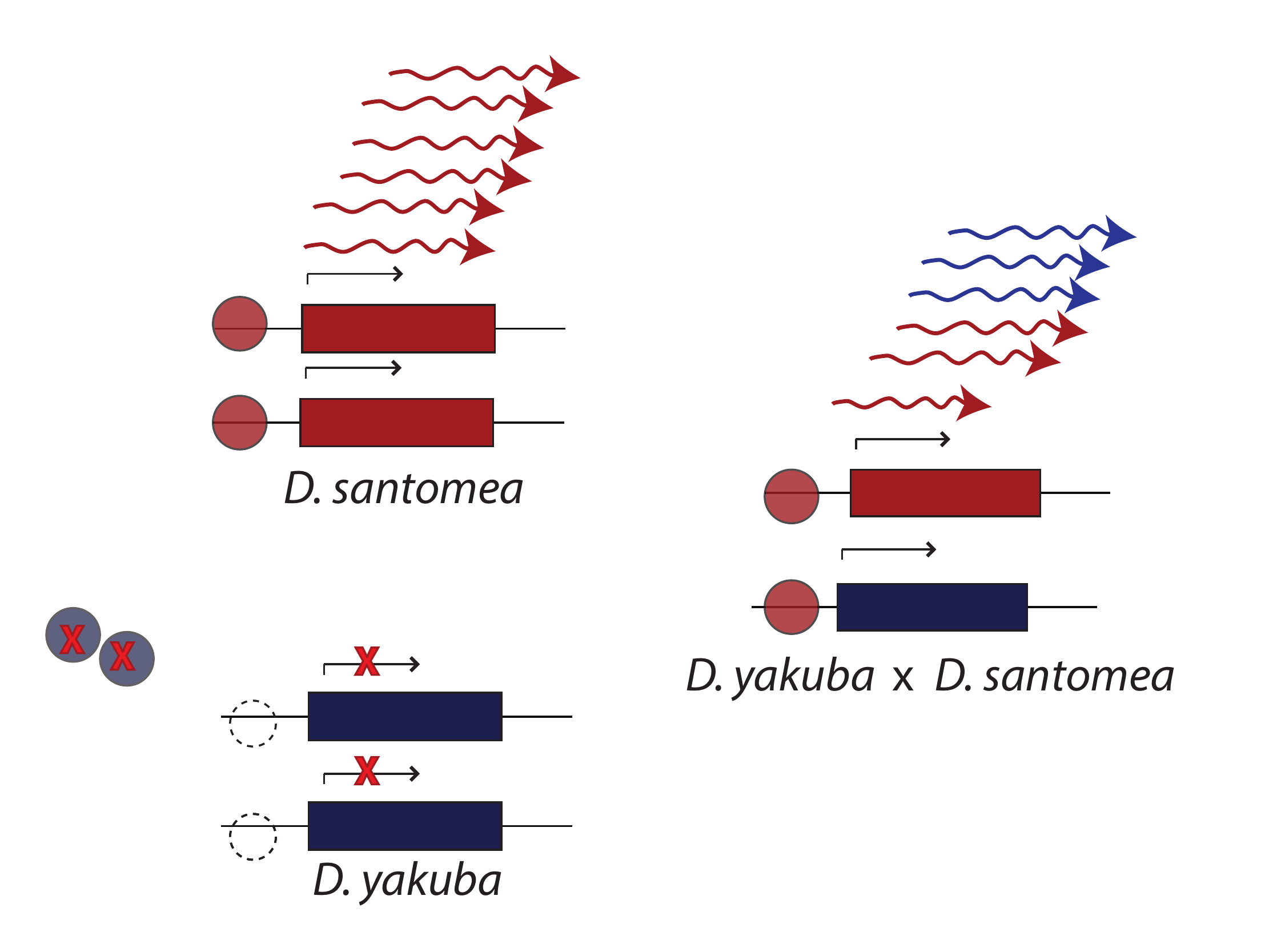}
     
    \caption{\label{AlleleActiveMechanism} Reactivation of alleles silenced \trans {} factors in hybrids. An allele in one species has been silenced due to inactivation of a \trans {} acting factor.  When hybridization adds active \trans {} factors, expression is driven from the locus. }
\end{figure}

New genes are not the only source of genetic novelty that appears in hybrid offspring.  Similar effects of allelic activation may be observed when transcripts are active in one species but silent in another (Figure \ref{AlleleActiveMechanism}).  A \emph{locus} previously silenced through inactivation of a \trans {} factor may be reactivated in the hybrid if the \cis {} regulatory module is intact.   We may also expect to observe creation of transcripts at \emph{loci} that were previously untranscribed if transcripts were newly formed in the active parent.   There are 77 total cases of allelic activation in \Dyak {} $\female \times$ \Dsant {} $\male$ crosses, 57 of these in the testes (Table \ref{AlleleActive}).   A total of  44 in \Dyak {} $\male \times$ \Dsant {} $\female$ crosses, 38 of which are in testes (Table \ref{AlleleActive}).  

The \Dsant {} allele is more likely to be activated by the \Dyak {} genome and/or cellular components for both types of crosses.  The prevalence of activation in the testes may be consistent with rapid divergence of testes-expressed alleles between species resulting in an excess of testes-specific activation. These \emph{loci} may be cases of reactivation at an allele that was historically silenced in \trans {} or activation of hybrid alleles after new gene formation in one parent (Figure \ref{AlleleActiveMechanism}).  The number of newly expressed alleles follows a fundamentally different pattern across tissues and direction of the cross compared with new gene formation likely reflecting the differing impacts of these two molecular mechanisms that form the underpinnings of regulatory changes in hybrids.

\subsubsection*{Regulatory changes in hybrids}

\begin{figure}
\begin{center}
\begin{subfigure}[b]{0.45\textwidth}
        \includegraphics[width=\textwidth]{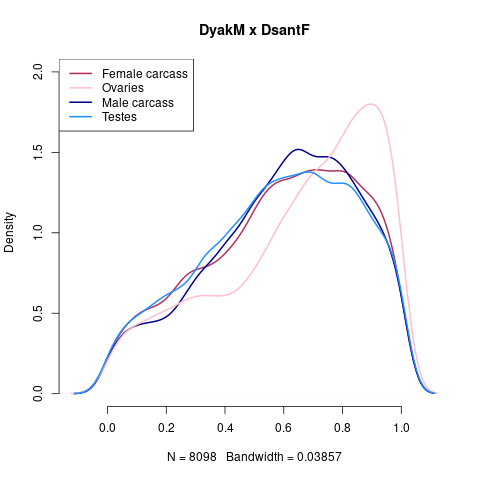}
            \end{subfigure}
            \begin{subfigure}[b]{.45\textwidth}
        \includegraphics[width=\textwidth]{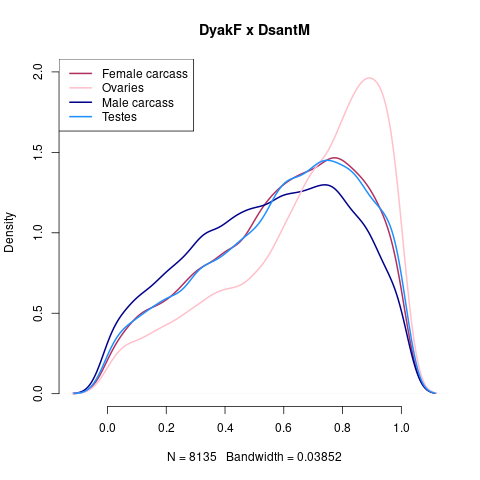}
            \end{subfigure}

\caption{Percent expression changes driven by \cis-regulatory changes in hybrid and reference strains for A) \Dsant {} $\female\times$\Dyak {} $\male$ and \Dyak {} $\female \times$ \Dsant {} $\male$.  Crosses in both directions show a skew toward \cis-regulatory variation that is most extreme in ovaries.\label{PercentCis}}
\end{center}
\end{figure}

It is already well established that quantitative regulatory changes occur in \emph{Drosophila} F$_1$ hybrids \cite{wittkopp2004, Ranz2004, Coolon2014} including \Dyak-\Dsant {}  crosses \cite{Llopart2012}.  We estimated \cis {} and \trans {} contributions to regulatory variation according to the methods used previously to assay \Dmel {} \cite{wittkopp2004,Ranz2004} and similar those for X-chromosome assays in \Dsant {} \cite{Llopart2012}.   We observe a strong correlation between ratios of expression for species-specific alleles in hybrids and the ratio of expression in the parental species (Spearman's $\rho >0.746$, $R^2>0.77$, $P<2.2 \times 10^-16$).  These results are consistent with constraint on expression variation among the \Dyak-\Dsant {} species complex. All tissues show a prevalence of \cis-regulatory effects (Figure \ref{PercentCis}).  However, ovaries show a more extreme skew toward \cis-regulatory divergence rather than \trans {} regulatory divergence than all other tissues.  These results may suggest tighter constraint on \trans {} acting factors that affect the ovaries, as these are directly related to fertility and offspring numbers.

\subsection*{Discussion}
\subsubsection*{New gene formation in hybrids} 
It is commonly assumed that DNA mutations underlie the genetic basis of evolutionary innovation.   Known sources of new gene formation include whole gene duplication, chimera formation, \emph{de novo} gene formation through new promoters, and TE induced activation.  These new mutations appear in populations and contribute genetic novelty that is subject to natural selection.   However, here we show that additional sources of new genes may appear without underlying changes to the DNA.  In this work we have shown that new transcripts can appear in hybrid \Dros.  Through interaction of \cis-regulatory regions and \trans {} acting factors, new genes can be created via epistatic interactions or through changes in chromatin modeling.  Hybrid crosses are partially sterile with greater mating resistance in the \Dsant {} females crossed with \Dyak {} males than the reciprocal cross \cite{coyne2004}.  Our work on new gene formation shows similar asymmetry in the creation of genetic novelty, as the number of newly expressed \emph{loci} depends on the direction of the cross.  

These results fundamentally challenge long-standing views in the field of genetic novelty.  Rather than requiring random mutations to appear in DNA, new genes can form immediately through introgression from a different genetic background.  The tempo of new gene formation under these scenarios is appears to be fundamentally different from the assumed molecular clock of new gene formation.   A ``burst" of new gene formation in F$_1$ hybrids results in multiple new transcripts appearing in a single generation.  Compared with estimates of duplicate gene formation in a clock-like progression of 8 per million generations \cite{RBH}, hybridization events can produce a greater number of new transcripts in a single generation than would normally be observed over millions of years.  \Dyak-\Dsant {} hybrids form in the wild at the introgression zone.  Hybrid flies from nature contain introgression tracts of 0-1\% per fly up to 5\% of the genome across the entire population \cite{Turissini2017}.  With regular hybridization and introgression in nature, we expect similar processes to contribute to new gene formation in the wild.  

\subsubsection*{Rapid evolution in the testes}
We observe the most extreme effects of allelic activation in the testes, regardless of the direction of the cross, consistent with previous work larger effects of sterility in the heterogametic sex \cite{presgraves2018}.  Such results are consistent with rapid expression divergence  in the testes and accessory glands \cite{dorus2008, Karr2019, Zhao2014,Assis2013, rogers2017}.  Historically silenced alleles through inactivation of \emph{trans} factors can produce regulatory variation between these two closely related species.  Reintroduction of functional \emph{trans} acting factors can then resurrect alleles that may have evolved under reduced functional constraints, accumulating divergence.   Activation of such alternative alleles may offer one contribution to differences in hybrids that is not typically observed in either parental species.    These transcripts that have accumulated divergence may be allowing a broader search of the sequence space near the previous optimum.  When habitats shift, this could be an important source of amino acid changes or even regulatory variation. Under the average scenario with static selective schemes, we would expect this variation to be maladaptive.  However, in rare cases this substrate of genetic novelty may contribute `hopeful monsters' of the genetic world that offer novelty in the face of selection.

\subsubsection*{Genome Dominance}
\begin{figure}
    \centering
    \begin{subfigure}[b]{0.45\textwidth}
        \includegraphics[width=\textwidth]{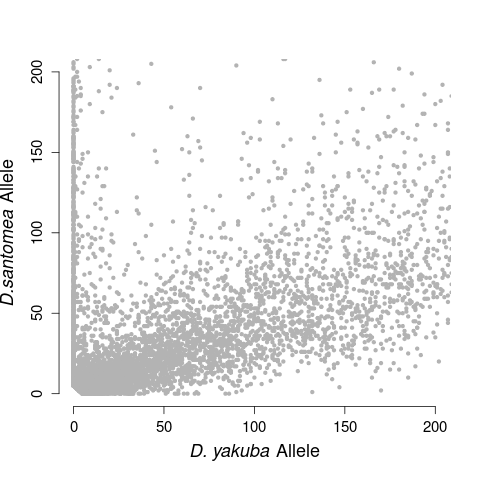}
           \end{subfigure}
    ~ 
    \begin{subfigure}[b]{0.45\textwidth}
        \includegraphics[width=\textwidth]{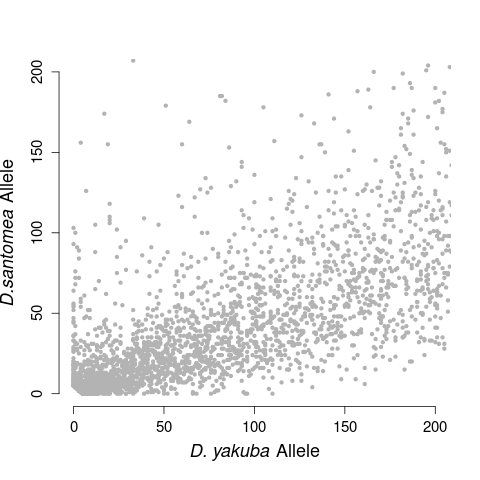}
          \end{subfigure}
        \caption{\label{ABalanceMain} Allelic balance in hybrid \Dros.  A)  \Dsant {} $\female\times$\Dyak {} $\male$ F$_1$ testes. B)  \Dsant {} $\female\times$\Dyak {} $\male$ F$_1$ ovaries. Testes sequences show a pattern of high expression of the \Dsant {} allele in hybrids with little or no expression of the \Dyak {} allele.  The results are indicative of the \Dyak {} regulatory machinery driving ectopic expression of the \Dsant {} genome. A subset of these create new transcripts where none existed before. }
\end{figure}

The number of transcripts that appear in hybrids depends on the sex of the \Dyak {} parent (Figure \ref{ABalanceMain}).  These results imply that the \Dyak {} regulatory machinery is more likely to activate \Dsant {} genetic material than the converse, even though new genes represent newly transcribed \emph{loci} not expressed in either parent.  Allelic activation shows a 1.9-2.5 fold excess of activation of the \Dsant {} allele compared to the \Dyak {} allele.     These new transcripts are derived from regions active in one species, but silent in another.  Similar cases of activation fit models of overdominant expression previously identified in \Dmel-\Dsech {} hybrids \cite{McManus2010}. 

The patterns of allelic activation across tissues are strikingly different from new gene formation.  While new gene formation depends on the direction of the cross, allelic activation is most common in the testes for each of the reciprocal cross.  Here, the mechanism that produces regulatory changes influences the number of \emph{loci}, the tissue specificity, and the tempo of allelic activation compared with new gene formation. Hence, the tempo and progression of evolution is expected to be entangled with molecular mechanisms driving innovation in genetic novelty.

\subsubsection*{The tempo of new gene formation}
Previous work on new gene formation has struggled to explain how protogenes might appear and then accumulate modifying mutations that would alter function.  The wait time for new genes to appear can be long if mutation rates are low \cite{Rogers2015}.   In contrast, new gene formation in hybrids does not require new mutations.  Rather, epistatic interactions between two genomes that have never met before will offer immediate sources of new transcripts.   A question then arises with respect to how these sequences might acquire appropriate translation signals, splice junctions, and new proteins.  One possibility is that these are non-coding RNAs.   Similar explanations have been suggested as a mechanism of action for \emph{de novo} genes \cite{carvunis2012,siepel2009,schlotterer2015}.

If new transcripts rely on epistatic interactions, they may not necessarily be permanent in genomes.  However, through introgression of appropriate \trans-acting factors, these may be more stably incorporated into the population.   Such introgression would not be observed necessarily at the site of the gene, but might sometimes affect the \trans {} factor.   This new gene formation would occur in a `burst' of transcripts, unlike the clock-like accumulation of single base pair changes or gene duplication.  These dynamics might mimic bursts of TE movement, except that the \emph{loci} are reproducible across replicates rather than variable products of random insertion.

\subsubsection*{Regulatory changes and hybrid phenotypes}
 These results may suggest that introgression could be a strong source of new genes when new genes are needed.   As with every case of new gene formation, most are expected to be detrimental or neutral.  These non-advantageous mutations are expected to be lost from populations quickly.   Yet, as a source of evolutionary change, they have the potential to alter gene content in hybrids in a way that may contribute to unique hybrid phenotypes \cite{Landry2007}, and potentially new hybrid species \cite{comeault2018}.   We observe a greater number of abnormally expressed genes in the germline, with a significant effect from the direction of the cross (Table 1, Table 3).

 \subsubsection*{Future Directions}
 For the moment, new gene formation and allelic activation have been surveyed in F$_1$ hybrids.  Several open questions remain.   How do these transcripts behave in F$_2$ hybrids and later generations after pulses of introgression?  Can we identify counterparts in natural populations where introgression is common?  How does new gene formation depend on genetic distance between species?  Can crosses within species produce similar cases of \emph{de novo} formation?   These questions all deserve further inquiry in future studies.   It is possible that their answers will offer new and different stories about the genetics of evolutionary innovation and creation of new genes in nature.

 \subsection*{Methods}
 \subsubsection*{Sample processing}
 To identify cases of new gene formation in hybrid \Dros, we performed stranded RNAseq on 4 replicates each of the \Dyak {} reference genome strain (Stock Number 14021-0261.01) and the \Dsant {} reference genome strain (STO-CAGO 1402-3; Stock Number 14021-0271.01) as well as reciprocal hybrid crosses of \Dyak {} $times$ \Dsant {}.  We established reciprocal crosses of the \Dyak {} and \Dsant {} reference strains.  Flies were reared on cornmeal and molasses medium at 23$^\circ$C in a temperature and humidity controlled incubator.  Adult males and females were dissected at 5-7 days post eclosion.   We separated testes plus accessory glands versus adult carcass for males and ovaries plus uterus versus adult carcass for females.  Each of 5 replicates consisting of 5 dissected individuals were flash frozen in liquid nitrogen and stored at -80$^\circ$C until extraction.  
 
RNA was extracted using Zymo DirectZol RNA microprep Kit without DNase.  RNA was converted to cDNA using the Illumina TruSeq mRNA HT Library Prep.  Libraries were prepared using 8M fragmentation and 15 cycles of amplification. We used secondary cleanup step with Ampure XP beads to remove lower molecular weight primer-dimers. Transcriptomes were sequenced on a single lane of an Illumina HiSeq4000 using the Illumina 150bp paired end (PE) Cluster kit at the Duke University sequencing core (DUGSIM).  One replicate of \Dyak {} female carcass had no reads due to barcode failure.  We proceeded with data for the 4 highest coverage replicates, keeping matched gonads and soma together.  Replicates show consistent patterns of support for reference alleles across batches (Figure \ref{BatchGonad}-\ref{BatchSoma}).
 
 Sequences were aligned to the genome using tophat-2.1.1 under default parameters using the \Dyak {} r.1.0.5 as a guide.  Transcripts were annotated and quantified in each replicate via cufflinks, with the \Dyak {} r.1.0.5 annotations as guides (-G dyak-all-no-analysis-r1.05.fmt.gff).  The resulting transcript annotations and \Dyak {} r.1.0.5 were merged using cuffmerge.  The new annotation file including hybrid transcripts was then used to quantify transcripts in cufflinks and to test for differences in expression using cuffdiff.  All other parameters were set to default with FPKM normalization.  Sequencing depth for RNAseq alignment files was calculated using samtools depth.
 
\subsubsection*{New gene formation}
 We identified putative new genes in $F_1$ hybrid offspring of \Dyak {} and \Dsant {} crosses.  We required that new genes have a minimum FPKM of 5 in each of 4 hybrid replicates and all \Dsant {} and \Dyak {} reference replicates have FPKM$<$1.0 (Table \ref{NewGenes}).   Transcripts matching known genes were removed from candidates of new gene formation and were classified as activation of existing genes in novel tissues (Table \ref{ActivatedGenes}.  Size selection in isoseq is biased against shortest fragments and the challenges of aligning short sequences in long read aligners may reduce confirmation rates.    
 
Reference genomes may influence mapping and thereby complicate identification of new transcripts.  To ensure that results of new gene formation are not driven solely by reference bias we performed similar analysis with the PacBio Tai18E2 \Dyak {} reference genome, a more contiguous but unannotated PacBio generated assembly.  We de novo assembled \Dyak {} and \Dsant {} transcriptomes for reference strains using the Oyster River Protocol and mapped assembled transcripts to the \Dsant {} assembly using minimap2.17 using options -a -x splice -u f --secondary=no -C 5.  These were converted to gtf files and merged using cdna\_cupcake from Sqanti3 v. 4.0.  A new gtf annotation file was created for each tissue to identify regions with transcripts.   We then aligned RNAseq data for each replicate with these new annotations for Tai18E2 using Tophat and quantified expression using cufflinks and cuffdiff as described above.     
 
\subsubsection*{Evolution in \cis {} and \trans}
To identify potential cases of allele specific expression or allele specific activation, we used samtools mpileup and bcftools consensus caller to identify SNPs that differentiate \Dyak {} and \Dsant {} transcripts in RNAseq files, emitting all sites.  The resulting VCF was used to quantify transcript support for each reference and non-reference SNP.    We identified allelic activation in each cross and tissue as sites with a minimum of 10 reads per site at differentiating SNPs in hybrids, with coverage less than 2 reads per site at differentiating sites in one of the two reference genomes.  All four replicates were required to display this pattern for gene to be candidates for allelic activation.  

Allele specific coverage depth data were aggregated across replicates for each cross and tissue type.  The \cis-\trans {} interactions were estimated according to methods similar to previous work \cite{Llopart2012, Coolon2014}.   We corrected estimates so that each replicate or tissue would have identical numbers of reads across all sites associated with allele specific expression.  We downsampled reference read counts in hybrid samples to adjust for genome wide reference bias using binomial sampling on the reference count.   We used Fisher's hypergeometric distribution to downsample the sample with the highest number of reads so that each sample in the comparison (H1/H2 vs P1/P2) had equal read counts.  The \cis regulatory changes $c$ were given by $\log_2(H1/H2)$ and \trans {} regulatory changes $t$ by $\log_2(P1/P2) - \log_2(H1/H2) $ according to previous work.  The percent \cis {} regulatory changes was given by $\frac{\vert c \vert}{\vert c \vert+\vert t \vert}$. 

 \clearpage
 
\subsubsection*{ Acknowledgements}
This work was funded by startup support from the University of North Carolina, Charlotte and by NIH NIGMS R35-GM133376 to RLR.  We thank the Duke University School of Medicine for the use of the Sequencing and Genomic Technologies Shared Resource, which provided Illumina sequencing.  Mt. Sinai Icahn School of Medicine sequencing core produced PacBio isoseq data.  Jon Halter, Michael Moseley, Chad DeWitt, Chuck Price, and Chris Maher provided help with software installation and functionality on the UNCC HPC system.

\subsubsection*{Data Availability}
Supplementary data are available in Dropbox during review at \url{https://www.dropbox.com/sh/sesujw92psc8s16/AAAb5LIPLVAlPHMqsH8Jtae0a?dl=0}. Supporting sequence data fastq files will be released from the SRA prior to publication.  Submission is in process under SRA SUB10213145, currently subject to unexpected system-wide delays in SRA sample processing (since August 2, 2021; continuing as of August 17,2021). 
 \clearpage

\bibliographystyle{PLoS}
\bibliography{HybridDrosophila}
\clearpage

\begin{table}
\center
\caption{New Genes formed in hybrids}
\begin{tabular}{llcc}

Offspring Sex &Tissue & \Dsant {} $\male \times$ \Dyak {} $\female$ & \Dyak {} $\male \times$  \Dsant {} $\female$   \\
\hline 
 \female & carcass & 0  &0 \\  
 & ovaries & 10 & 0\\
\male & carcass & 0 & 1 \\
  &  testes & 2 & 18 \\
\hline
\end{tabular}
\\
mean ref $<$ 1, mean hyb $>$ 5, min Hyb$>$ max Ref\\

Fisher's Exact Test $P=1.488\times10^{-6}$\\
\label{NewGenes}
\end{table}

\clearpage

\begin{table}
\center
\caption{Newly activated alleles in hybrid \Dros}
\begin{tabular}{lclll}
Cross & Offspring Sex & Tissue & Sant active & Yak active \\
\hline
\Dyak {} $\male$ $\times$ \Dsant {} $\female$ & $\male$ & testes & 27 & 11\\
    	&  & caracass & 1 & 3 \\
	& $\female$& ovaries & 2 & 0 \\
	& & carcass & 0 & 0 \\
\hline
\Dyak {} $\female$ $\times$ \Dsant {} $\male$ & $\male$&  testes & 37 & 20\\
    	&  &carcass & 4 & 3 \\
	& $\female$ &ovaries & 3 & 1 \\
	& & carcass & 8 & 1 \\
\hline 

\end{tabular}
\label{AlleleActive}
\end{table}

\clearpage

\clearpage

\clearpage

\clearpage

\clearpage
\clearpage


\subsection*{Supporting Information}

\pagenumbering{arabic}
\renewcommand{\thefigure}{S\arabic{figure}}
\renewcommand{\thetable}{S\arabic{table}}
\setcounter{figure}{0}
\setcounter{table}{0}

\clearpage

\begin{table}
\center
\caption{Known genes activated in hybrids}
\begin{tabular}{llcc}

Offspring Sex &Tissue & \Dsant {} $\male \times$ \Dyak {} $\female$ & \Dyak {} $\male \times$  \Dsant {} $\female$   \\
\hline 
 \female & carcass & 1  &0 \\  
 & ovaries & 0 & 1\\
\male & carcass & 16 & 0 \\
  &  testes & 0 & 6 \\
\hline
\end{tabular}
\\
mean ref $<$ 1, mean hyb $>$ 5, min Hyb$>$ max Ref\\

Fisher's Exact Test $P=5.779\times10^{-6}$\\
\label{ActivatedGenes}
\end{table}
\clearpage
\begin{table}
\center
\caption{Correlation between hybrid and reference expression ratios at autosomal genes}
\footnotesize
\begin{tabular}{llccccc}
Cross & tissue & Spearman's $\rho$ & P-value & $R^2$  & P-value \\
\hline
\Dyak {} $\female$ $\times$ \Dsant {} $\male$ & mcarcass  & 0.7461  & $< 2.2\times10^{-16}$ &  0.885 & $< 2.2\times10^{-16}$ \\
	& fcarcass & 0.7612 & $< 2.2\times10^{-16}$ & 0.8511 & $< 2.2\times10^{-16}$ \\
	& testes  & 0.8534 & $< 2.2\times10^{-16}$   &  0.8045 & $< 2.2\times10^{-16}$ \\
	& ovaries & 0.8404 &  $< 2.2\times10^{-16}$ & 0.7768 &  $< 2.2\times10^{-16}$ \\
\hline
\Dyak {} $\male$ $\times$ \Dsant {} $\female$& mcarcass & 0.7486 & $< 2.2\times10^{-16}$ & 0.8613 & $< 2.2\times10^{-16}$ \\
 & fcarcass & 0.7491  & $< 2.2\times10^{-16}$  &0.8354 & $< 2.2\times10^{-16}$ \\
  & testes & 0.7984  &  $<2.2\times10^{-16}$& 0.8679 & $< 2.2\times10^{-16}$ \\
   & ovaries & 0.8044&  $<2.2\times10^{-16}$& 0.7843&$< 2.2\times10^{-16}$\\
   \hline
\end{tabular}
\end{table}

\clearpage

\begin{figure}
    \centering
    \begin{subfigure}[b]{0.23\textwidth}
        \includegraphics[width=\textwidth]{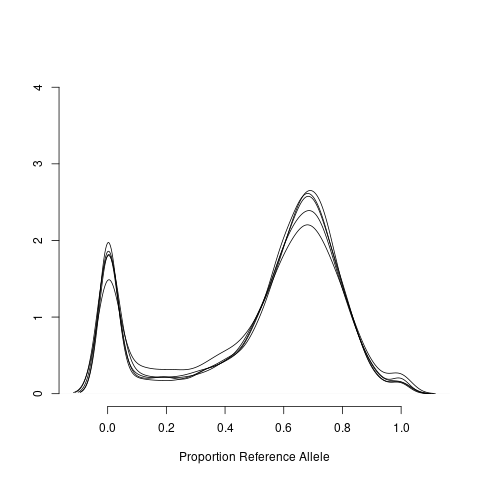}
           \end{subfigure}
    ~ 
    \begin{subfigure}[b]{0.23\textwidth}
        \includegraphics[width=\textwidth]{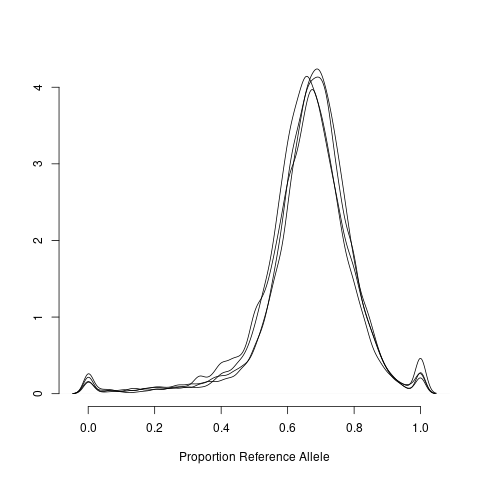}
          \end{subfigure}
    ~ 
    \begin{subfigure}[b]{0.23\textwidth}
        \includegraphics[width=\textwidth]{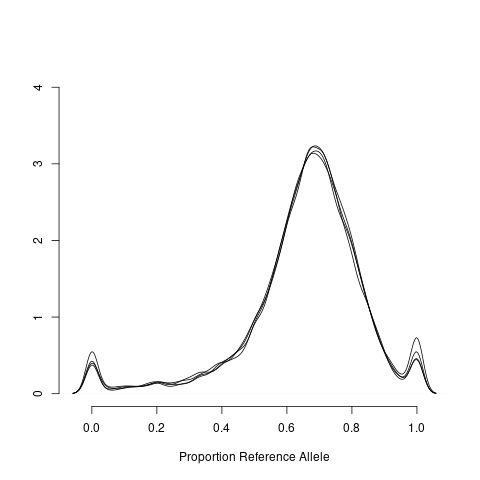}
      
    \end{subfigure}
      \begin{subfigure}[b]{0.23\textwidth}
        \includegraphics[width=\textwidth]{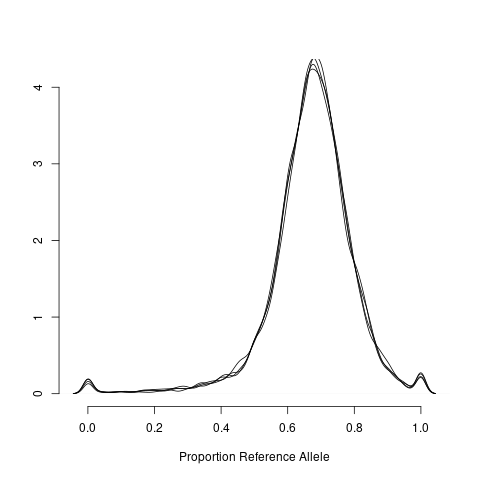}
      
    \end{subfigure}
    \caption{\label{BatchGonad} Density plots showing allelic balance in \Dyak$\times$\Dsant {} hybrid \Dros.  A)  \Dsant {} $\female\times$\Dyak {} $\male$ F$_1$ testes. B)  \Dsant {} $\female \times$ \Dyak {}  $\male$ F$_1$ ovaries.   C)  \Dyak {} $\female\times$\Dsant {} $\male$ F$_1$ testes.   D)  \Dyak {} $\female\times$\Dsant {} $\male$ F$_1$ ovaries.  All replicates show a bias toward expression of the \Dyak {} allele over the \Dsant {} allele. }
\end{figure}
\clearpage

\begin{figure}
    \centering
    \begin{subfigure}[b]{0.23\textwidth}
        \includegraphics[width=\textwidth]{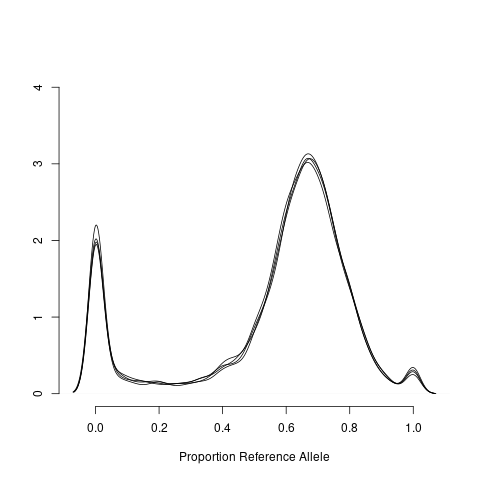}
           \end{subfigure}
    ~ 
    \begin{subfigure}[b]{0.23\textwidth}
        \includegraphics[width=\textwidth]{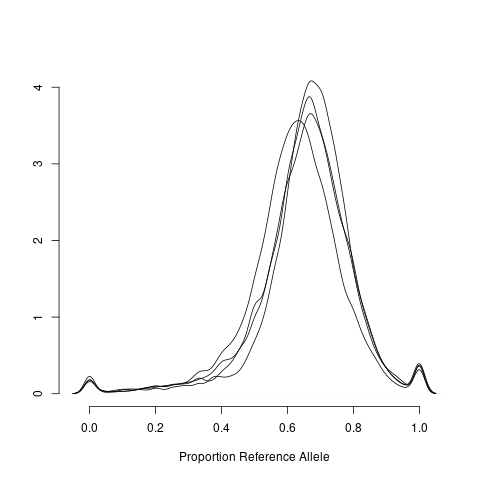}
          \end{subfigure}
    ~ 
    \begin{subfigure}[b]{0.23\textwidth}
        \includegraphics[width=\textwidth]{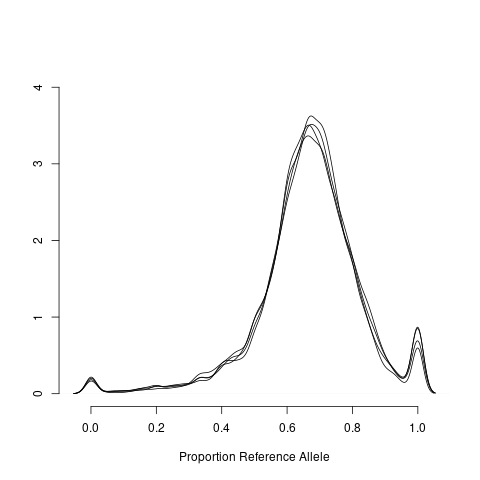}
      
    \end{subfigure}
      \begin{subfigure}[b]{0.23\textwidth}
        \includegraphics[width=\textwidth]{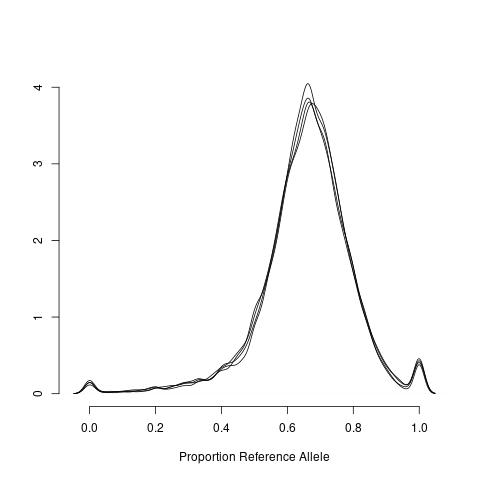}
      
    \end{subfigure}
    \caption{\label{BatchSoma}Density plots showing allelic balance in \Dyak$\times$\Dsant {} hybrid \Dros {} soma.  A)  \Dsant {} $\female\times$\Dyak {} $\male$ F$_1$ male carcass. B)  \Dsant {} $\female\times$\Dyak {} $\male$ F$_1$ female carcass.   C)  \Dyak {} $\female\times$\Dsant {} $\male$ F$_1$ male carcass.   D)  \Dyak {} $\female\times$\Dsant {} $\male$ F$_1$ female carcass.  All replicates show a bias toward expression of the \Dyak {} allele over the \Dsant {} allele.   }
\end{figure}
\clearpage

\begin{figure}
    \centering
    \begin{subfigure}[b]{0.23\textwidth}
        \includegraphics[width=\textwidth]{Figures/ExpCorrMxF-testes.png}
           \end{subfigure}
    ~ 
    \begin{subfigure}[b]{0.23\textwidth}
        \includegraphics[width=\textwidth]{Figures/ExpCorrMxF-ovaries.png}
          \end{subfigure}
    ~ 
    \begin{subfigure}[b]{0.23\textwidth}
        \includegraphics[width=\textwidth]{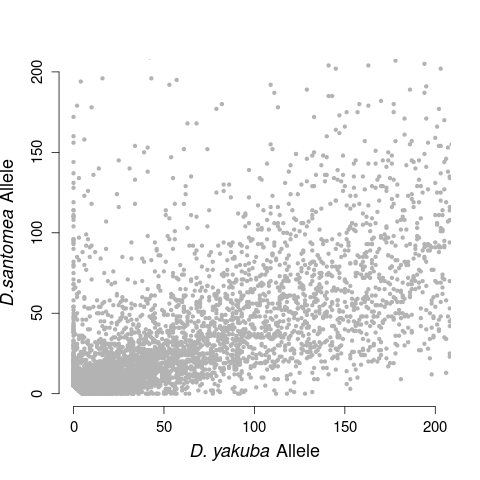}
      
    \end{subfigure}
      \begin{subfigure}[b]{0.23\textwidth}
        \includegraphics[width=\textwidth]{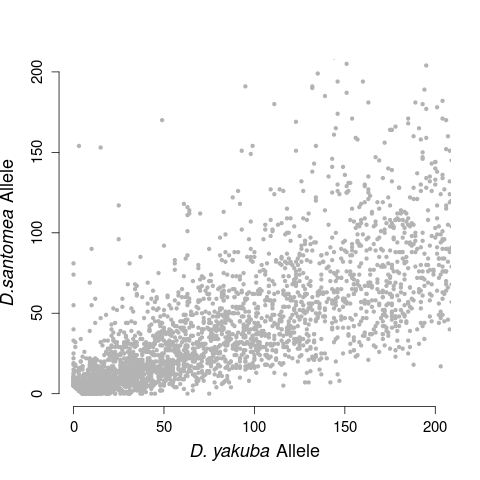}
      
    \end{subfigure}
    \caption{\label{ABalance} Allelic balance in \Dyak$\times$\Dsant {} hybrid \Dros.  A)  \Dsant {} $\female\times$\Dyak {} $\male$ F$_1$ testes. B)  \Dsant {} $\female\times$\Dyak {} $\male$ F$_1$ ovaries.   C)  \Dyak {} $\female\times$\Dsant {} $\male$ F$_1$ testes.   D)  \Dyak {} $\female\times$\Dsant {} $\male$ F$_1$ ovaries.  All replicates show a bias toward expression of the \Dyak {} allele over the \Dsant {} allele.  Some sites show silencing for the \Dyak {} allele but high expression for the \Dsant {} allele, reflecting high activation of the \Dsant {} sequence.  Some of these are new genes in hybrid \Dros.   Axes are truncated for visualization.  }
\end{figure}
\clearpage

\begin{figure}
\begin{center}
\begin{subfigure}[b]{0.4\textwidth}
        \includegraphics[width=\textwidth]{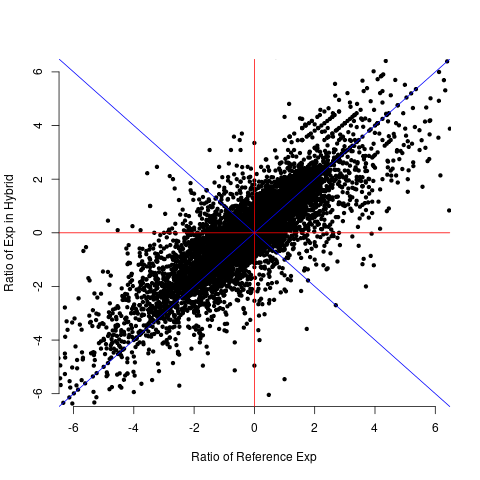}
            \end{subfigure}
            \begin{subfigure}[b]{.4\textwidth}
        \includegraphics[width=\textwidth]{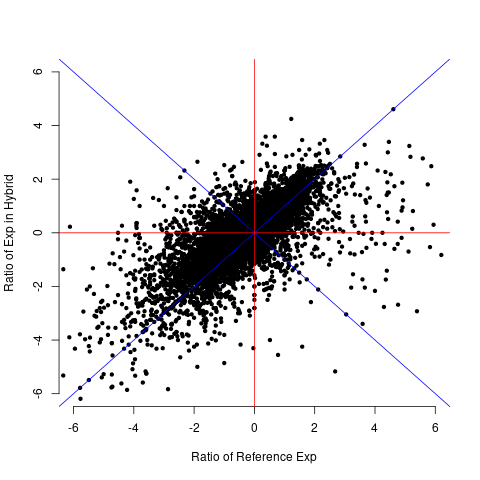}
            \end{subfigure}

 \begin{subfigure}[b]{.4\textwidth}
        \includegraphics[width=\textwidth]{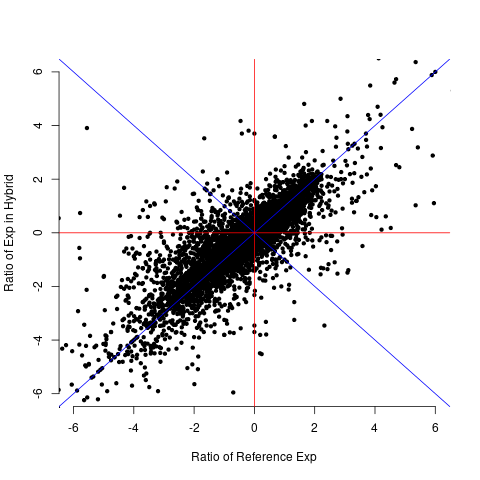}
            \end{subfigure}
             \begin{subfigure}[b]{0.4\textwidth}
        \includegraphics[width=\textwidth]{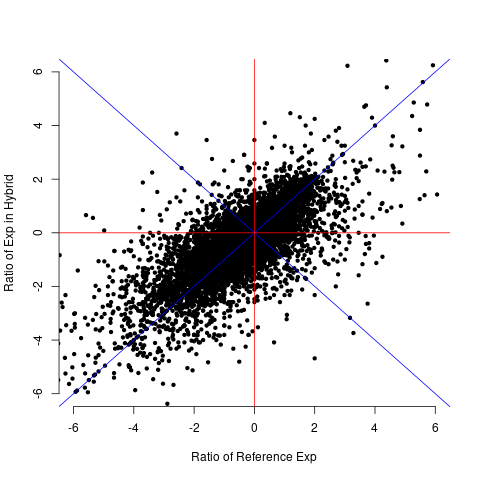}
            \end{subfigure}

\caption{Normalized allele specific expression in hybrid and reference strains for \Dyak {} $\female\times$\Dsant {} $\male$. A) Testes B) Male Carcass C) Ovaries  D) Female Carcass.  Ratios of expression in reference strains predict expression in hybrids well for most genes. Gonads show tighter correllation than carcass, which contains more diverse sets of tissues. }
\end{center}
\end{figure}
\clearpage

\begin{figure}
\begin{center}
\begin{subfigure}[b]{0.4\textwidth}
        \includegraphics[width=\textwidth]{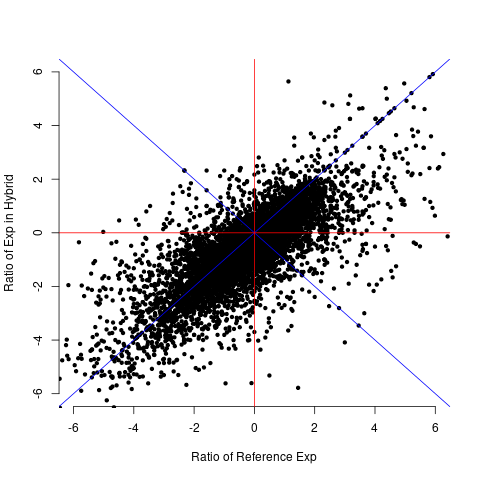}
            \end{subfigure}
            \begin{subfigure}[b]{.4\textwidth}
        \includegraphics[width=\textwidth]{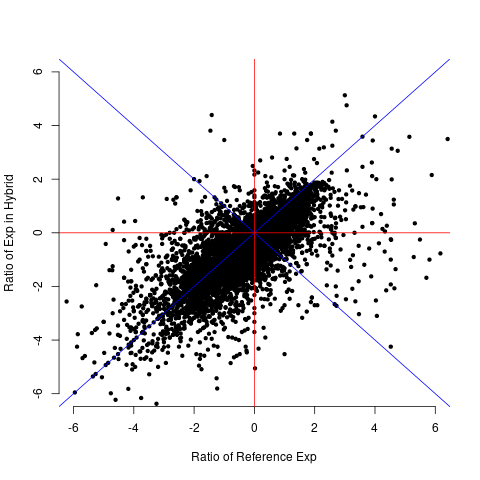}
            \end{subfigure}

 \begin{subfigure}[b]{.4\textwidth}
        \includegraphics[width=\textwidth]{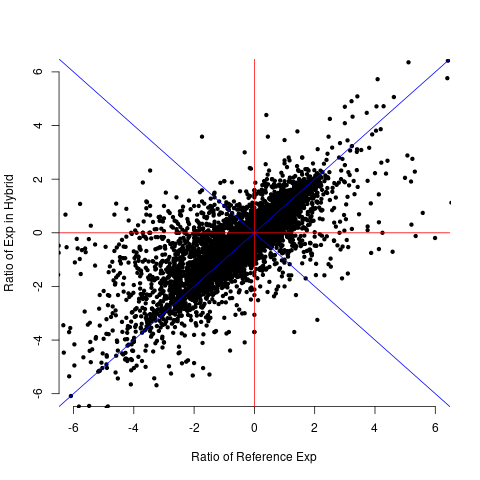}
            \end{subfigure}
             \begin{subfigure}[b]{0.4\textwidth}
        \includegraphics[width=\textwidth]{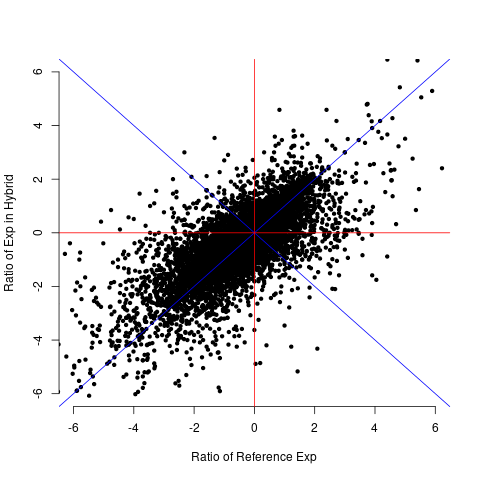}
            \end{subfigure}

\caption{Normalized allele specific expression in hybrid and reference strains for \Dsant {} $\female\times$\Dyak {} $\male$ for A) Testes B) Male Carcass C) Ovaries  D) Female Carcass.  Ratios of expression in reference strains predict expression in hybrids well for most genes. Gonads show tighter correllation than carcass, which contains more diverse sets of tissues. }
\end{center}
\end{figure}

%
%
%

\clearpage

\begin{figure}
\begin{center}
\includegraphics[width=.8\textwidth]{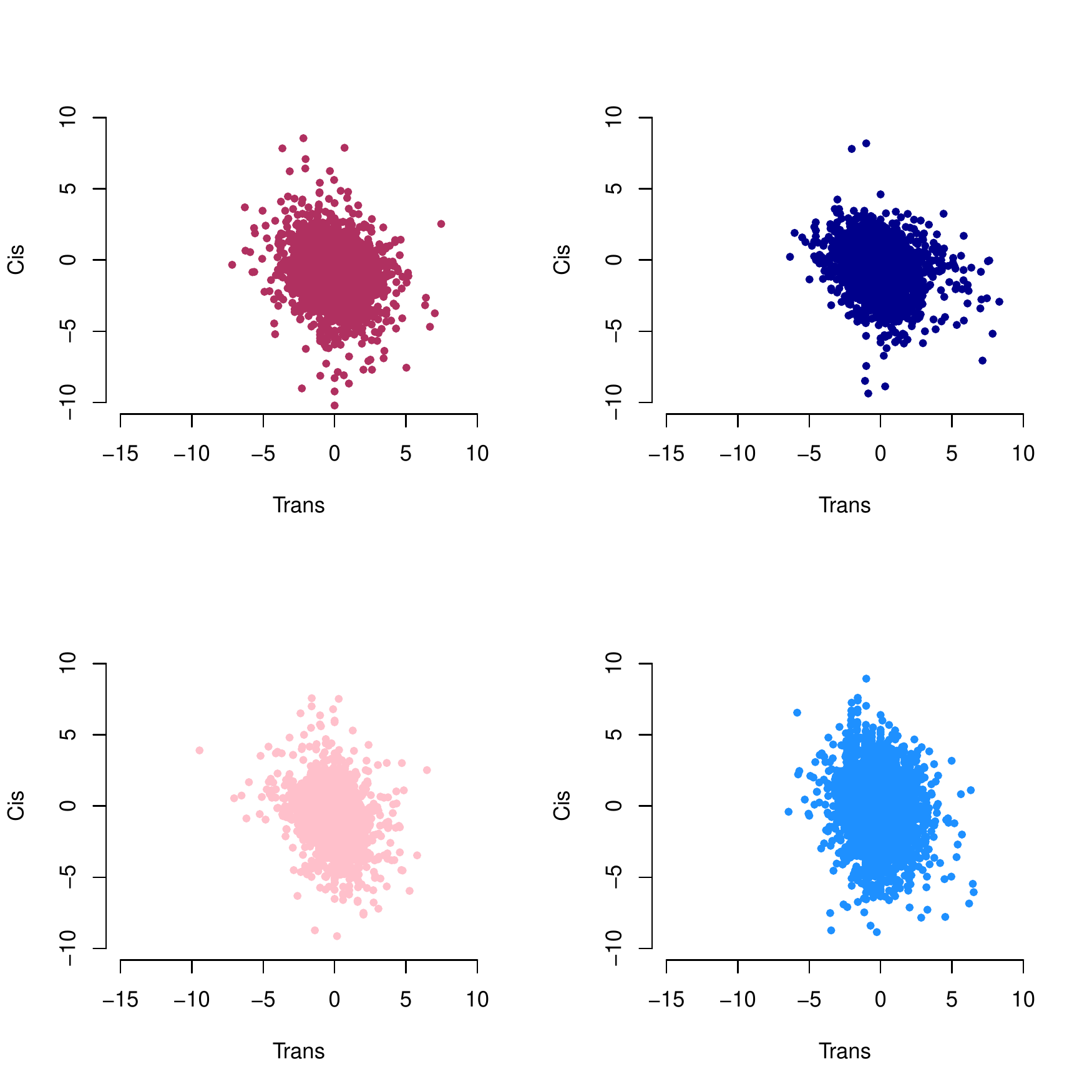}
\caption{\cis {} and \trans {} contributions to regulatory changes for autosomes in offspring of \Dyak {} $\female\times$\Dsant {} $\male$.}
\end{center}
\end{figure}
\clearpage

\begin{figure}
\begin{center}
\includegraphics[width=.8\textwidth]{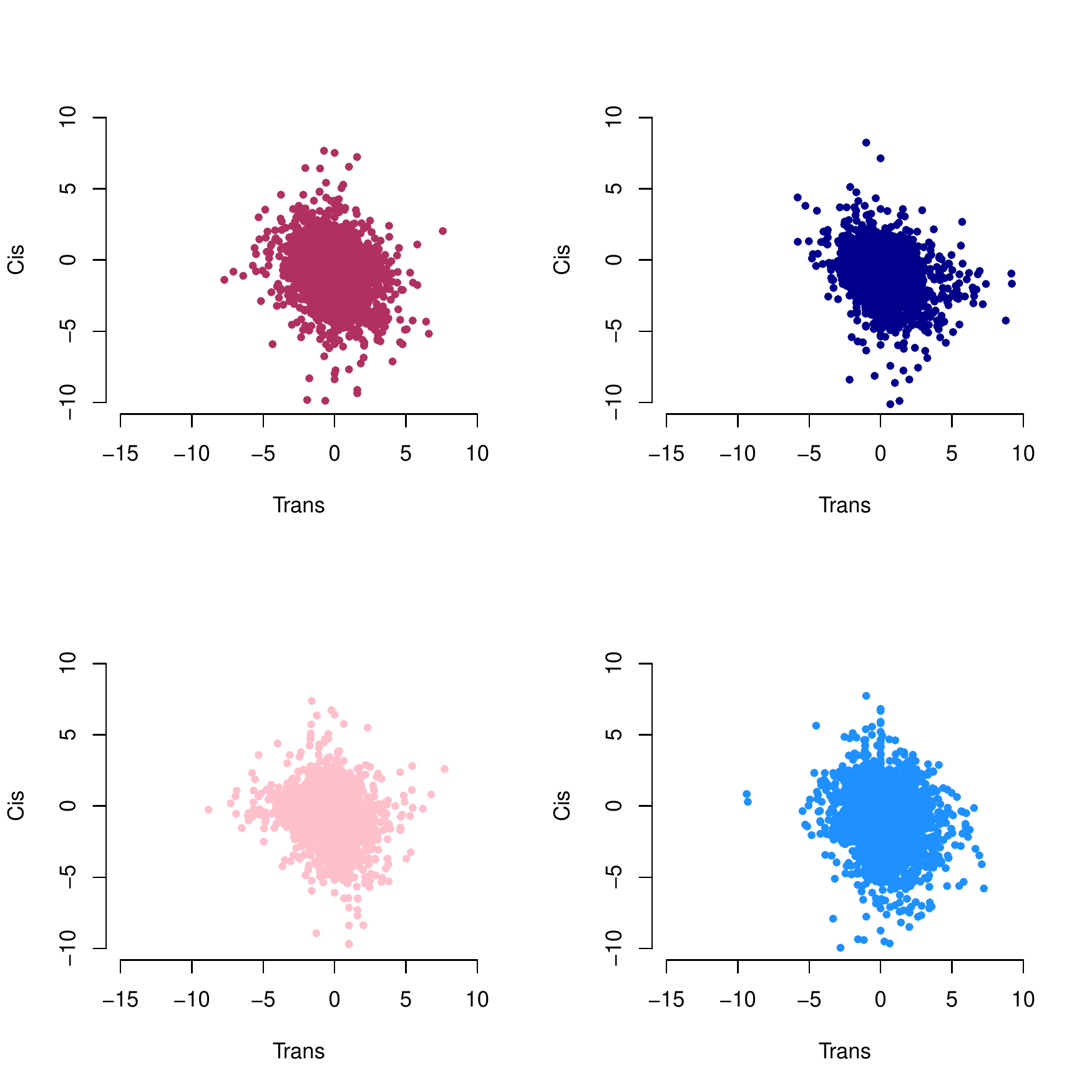}
\caption{\cis {} and \trans {} contributions to regulatory changes for autosomes in offspring of \Dyak {} $\female\times$\Dsant {} $\male$.}
\end{center}
\end{figure}
\clearpage

\end{document}